\documentclass[a4paper,12pt]{article}
\usepackage{amsmath,amsthm,amssymb}

\begin{document}
\begin{titlepage}
\begin{flushright}
\begin{tabular}{l}

UTHEP-568 \\
 2008 July
\end{tabular}
\end{flushright}

\vspace{5mm}

\begin{center}
{\Large \bf AdS/QCD and Type 0 String Theory}
\baselineskip=24pt

\vspace{20mm}

{\large Takashi Saitou\footnote{e-mail:saitout@het.ph.tsukuba.ac.jp}}\\
{ Institute of Physics, University of Tsukuba, \\
Tsukuba, Ibaraki 305-8571, Japan} \\

\vspace{20mm}

\end{center}

\begin{abstract}
We study gauge/string duality 
with flavor degrees of freedom 
in type 0B string theory.
We construct a model by  placing the $N_f$ probe $D9_-$/anti-$D9_-$ brane pairs 
in the background of the $N_c$ electric $D3$-branes.
This model has some  features in common with QCD, as in AdS/QCD model.
By analyzing the effective action, we obtain the linear confinement behavior for the masses of  
the highly excited spin-1 mesons. 
\end{abstract}

\end{titlepage}
\newpage


\section{Introduction}

Gauge/string duality 
\cite{Maldacena:1997re}
 has provided us with new tools for
investigating  strongly coupled gauge theories.
Now, there are many  attempts to use  this duality  to construct 
holographic dual of  four-dimensional confining gauge  theories, 
such as QCD.
Basically, there are two approaches to 
construct holographic dual description of QCD, 
although the exact form has not yet been found. 

One is the top-down approach in which one starts from string theory.
A key point to obtain the holographic dual of 
QCD-like gauge theories in this approach 
is to add flavor degrees of freedom
in non-supersymmetric backgrounds created by $N_c$ color branes.
This
can be done by introducing $N_f$ flavor branes in the probe approximation 
$N_f \ll N_c$,
where the back reaction of the flavor branes can be neglected
\cite{Karch:2002sh}.
This has been applied to various supergravity models, for example
\cite{Kruczenski:2003uq}
\cite{Sakai:2004cn}.
These models have been successful in capturing many qualitative features of low energy physics of QCD. 
For a recent review on this subject, see \cite{Erdmenger:2007cm}.
 
Another approach, the bottom up approach, is to start from QCD and 
attempt to find the holographic dual theory by matching it to some properties of QCD.
This approach is often called AdS/QCD.
In the simplest model
\cite{Erlich:2005qh}
\cite{Da Rold:2005zs}, 
the five-dimensional curved spacetime is taken to be $AdS_5$ with a hard IR cutoff, as 
first introduced in 
\cite{Polchinski:2001tt}.
This IR cutoff can incorporate the quark confinement.
Chiral symmetry breaking is modeled by a classical solution of scalar fields in 
a gravitational background.
The UV conformal invariance is incorporated by the conformal isometry of $AdS$ space.
A modification of this background has been considered in order to reproduce 
the linear confinement behavior 
for the masses squared of highly excited mesons,   
i.e. $m^2_{n,S} \propto S+n$,
where  $S$ is spin and  $n$ is radial excitation number
\cite{Karch:2006pv}.

In this paper, 
we would like to consider a string theory set up, keeping  these nice properties of AdS/QCD model. 
High energy behavior of the known top down models is 
completely different from QCD since these models have been constructed by breaking 
conformal symmetry and supersymmetry.
Therefore, we need to find a non-supersymmetric string theory set up.
Type 0 string theory
\cite{Dixon:1986iz} can be considered as a natural candidate because 
there is no supersymmetry but there exists fermionic degrees of freedom in the open string spectrum 
stretched between different types of $D$-branes.\footnote{ 
As we will explain later, there is a doubled set of Ramond-Ramond fields.
Therefore, there are two different  types of $D$-branes, which are 
called $D_{\pm}$-branes.}$^{,}$\footnote{A relevance between the spectrum of QCD and
that of type 0 string theory has been discussed in 
\cite{Polyakov:1998ju}
\cite{Cotrone:2007gs}.}
Gauge/string duality has been considered in this theory
\cite{Klebanov:1998yya}. 
The authors have considered a configuration of
$N_c$  $D3_+$ branes in  type 0B string theory. 
The dual field theory has been conjectured to be $U(N_c)$ four-dimensional Yang-Mills theory 
coupled to  6 
adjoint scalar fields.
The gravitational background corresponding to the $D3$ branes has been found in 
\cite{Klebanov:1998yy}
\cite{Minahan:1998tm}.
This background has the properties that we need. IR cutoff is naturally introduced 
through the  logarithmic behavior of the dilaton background.
In the UV region, the background becomes AdS space asymptotically.

We  wish to study the chiral symmetry breaking and the linear confinement behavior
for meson masses
by adding flavor degrees of freedom to this background.
A  general framework realizing chiral symmetry breaking 
has been formulated in  
\cite{Casero:2007ae}.\footnote{This method has been applied to the Sakai-Sugimoto model in
\cite{Dhar:2007bz}
\cite{Bergman:2007pm}.}
The authors have considered a  system of intersecting $D3$-$D9$-$\overline{D9}$ branes.
It has been argued that the open string  tachyons stretched between $D9$ pairs 
can be used to describe the chiral symmetry breaking of a gauge theory.
These fields correspond to the scalar fields modeling the 
chiral order parameter in AdS/QCD model.
It has been also shown that the linear confinement behavior for highly excited spin-1 meson masses 
are automatic in the formalism if the IR background satisfy some appropriate conditions.

In this paper, we study a configuration of 
$D3_+$-brane and $D9_-/\overline{D9}_-$ probe brane system  in  type 0B string theory
using this framework. 
It will be  shown that the low energy world volume theory of $D3_+$ branes is
the $U(N_c)$ Yang-Mills gauge theory with 6 adjoint scalars 
coupled to $N_f$ fermions in the fundamental representation of $U(N_c)$.
The fermions belong to the fundamental representation of the $U(N_f)_L \times U(N_f)_R$ flavor symmetry. 
Based on the framework, we discuss the chiral symmetry breaking and the linear confinement. 
We show that the type 0 IR background can be made to satisfy the conditions for linear confinement behavior 
by choosing the parameters associated with the IR background.
Thus the background can be considered as  a concrete  example of the background
discussed in \cite{Casero:2007ae}.

The organization of this paper is as follows.
In section 2, we briefly review some basic features of type 0 string theories.
 Then, we introduce the $D3_+$-$D9_-/\overline{D9}_-$ brane configuration 
 and explain the open string spectrum.
In section 3, we present the gravitational background and effective action for mesons.
We study the solution of tachyon equation of motion  derived from the action in section 4.
In section 5, we show that 
the mass spectra of highly excited vector and axial vector mesons have the 
behavior of the linear confinement.
Section 6 is devoted to conclusions.


\section{The set up in type 0 string theory}

\subsection{Type 0 string theory}
Here we briefly review  facts about type 0 string theories.

Let us first see that the closed string spectrum of  type 0 string theories
\cite{Dixon:1986iz}.
The GSO projection acts on the left and right-moving sectors together
as $(1+(-1)^{F+\tilde{F}})/2$ in the NS-NS sectors and   $(1\pm(-1)^{F+\tilde{F}})/2$ 
in the R-R sectors, where plus-minus sign corresponds to 0B and 0A theories, respectively.
As a consequence of this GSO projection, 
type 0 string theories possess no spacetime supersymmetry.
Following the notation of 
\cite{Polchinski:1998},
 the closed string spectrum of  type 0A and 0B theories are summarized as follows : 
\begin{eqnarray}
\begin{array}{ccc}
{\rm type \hspace{5pt} 0A} : & (NS-,NS-) \oplus (NS+,NS+) \oplus (R+,R-) \oplus (R-,R+),  \\
{\rm type \hspace{5pt} 0B} : & (NS-,NS-) \oplus (NS+,NS+) \oplus (R+,R+) \oplus (R-,R-). 
\label{eq:closedspectrum}
\end{array} 
\end{eqnarray}
Both of these theories have no fermions in their spectra.
The massless bosonic fields are  same as these in the corresponding type II theory (A or B), 
but there are twice as many   Ramond-Ramond fields. 
Type 0 theories also contain a tachyon coming from the $(NS-,NS-)$ sector.

The existence of a doubled set of R-R fields implies  
that of two different kinds of $D$-branes and  corresponding anti $D$-branes. 
Let us  denote the two kinds of $(p+1)$-form gauge potentials as
$C_{p+1}$ and $\bar{C}_{p+1}$ corresponding 
to the R-R sectors in (\ref{eq:closedspectrum}), 
for each $p$. ($p$ is even(odd) 
in  0A(0B) theory.)
Then, we define the gauge potentials as
\begin{eqnarray}
C_{p+1,\pm} \equiv \frac{1}{\sqrt{2}}( C_{p+1} +\bar{C}_{p+1}).
\end{eqnarray}
$D$-branes which are charged with respect to $C_{\pm}$ are
called  $D_\pm$ brane, respectively.
In the case of $D3$ brane in  0B theory, 
the self-dual field strength $F_5 =d C_4$ and the anti-self-dual field strength
$\bar{F}_5= d\bar{C}_4 $ can be combined to form an unconstrained 5-form field strength.
$D3_\pm$ branes are called "electric" and "magnetic" branes, 
because of the Hodge duality transformation property $*F_{5\pm}=F_{5\mp}$
\cite{Klebanov:1998yya}.


\subsection{The D-brane configuration}
The D-brane configuration we consider consists of 
$N_c$ electric  $D3$-branes and  $N_f$ $D9_-/\overline{D9}_-$ brane pairs.
The massless open string spectrum for this configuration is given in
\cite{Polyakov:1998ju}
\cite{Bergman:1997rf}
\cite{Costa:1999qx}.
The results are summarized as follows\footnote{
$U(N_c)$ is the gauge group associated with the $D3$-branes
 and $U(N_f)_L (U(N_f)_R)$ is the group associated with the $D9$-branes(anti $D9$-branes).
 We denote open strings with one end attached to the $Dp_\pm$-brane and the
other end to the $Dq_\pm$-brane as $(p_\pm, q_\pm)$.
We also denote the strings which belong to the bifundamental representation of 
 $U(N)$ $\otimes$ $U(N')$ 
as $(N,N')$.}  
: 
\begin{itemize}
\item $(3_+, 3_+)$ strings : a gauge boson $A_\mu $ $(\mu =0,1,2,3)$ 
and six adjoint scalar fields $\phi^i$ $(i=1,\cdots,6)$ in the adjoint representation of $U(N_c)$ gauge group
\item $(3_+, 9_-)$ strings : a Weyl fermion $q$ in the  representation $(N_c,N_f)$
of $U(N_c) \times U(N_f)_L$  
\item $(3_+, \bar{9}_-)$ strings : a Weyl fermion $q $ in the representation 
$(N_c,N_f)$ of $U(N_c) \times U(N_f)_R$
\item $(9_-,9_-)$ strings : a gauge boson $A^L_M $ $(M=0,\cdots,9)$ in the adjoint representation of $U(N_f)_L$
\item $(\bar{9}_-, \bar{9}_-)$ strings : a gauge boson $A^R_M$ in the adjoint representation of  $U(N_f)_R$
\item $(9_-, \bar{9}_-)$ strings : a tachyon $T$ in the representation $(N_f,N_f)$
of $U(N_f)_L \times U(N_f)_R$
\end{itemize}
The massless modes of $(3_+, 3_+)$ strings  consist of $A_\mu$ and $\phi^i$ which  
belong to the adjoint representation of the gauge group $U(N_c)$.\footnote{
The mass terms of the scalar fields $\phi^i$ are in general produced via loop-corrections.
Thus, 
the low energy world volume theory of the $N_c$ electric $D3$-branes is 
expected to be pure Yang-Mills theory.}
From the $(3_+,9_-)$ and $(3_+,\bar{9}_-)$ strings, we obtain 
$N_f$ Weyl fermions $q$, which belong to the fundamental representation of the 
$U(N_c)$ gauge group.  
The chirality of $(3_+,9_-)$ fermions  is opposite to that of $(3_+,\bar{9}_-)$ ones. 
Therefore, the $U(N_f) \times U(N_f)$ gauge symmetry of the 
$N_f$ $D9_-/\overline{D9}_-$ pairs can be  interpreted as the $U(N_f)_L \times U(N_f)_R$ 
chiral symmetry.

In summary, the world volume effective theory on $N_c$ electric $D3$ branes of the above intersecting $D$-brane
system is  four-dimensional $U(N_c)$ Yang-Mills theory with 6 adjoint scalars 
coupled to $N_f$ fermions with $U(N_f)_L\times U(N_f)_R$
chiral symmetry.


\section{The Holographic dual model}
Now let us describe the holographic dual of the above mentioned gauge theory.
We first introduce 
the background which corresponds to the electric $D3$ brane solution of type 0B string theory.
Then, we give the meson effective action.
There exist KK modes coming from $S^5$.
These degrees of freedom are generically present in a string theory set up.
We are not interested in these KK modes because these modes do not appear in QCD.
Thus, we ignore these modes as in other top-down approaches.
For this, we assume that meson fields are independent of the coordinate of $S^5$.
We can obtain the 5D effective action for mesons 
by the dimensional reduction of the action.


\subsection{The background}
The low energy effective action of type 0B string theory is given in
\cite{Klebanov:1998yya}, as
\begin{eqnarray}
&& S  = \int d^{10}x\Big[ e^{-2\phi}
\Big(R+4(\partial_M \phi)^2
-\frac{1}{4}(\partial_M t)^2-\frac{1}{4}g(t) 
\Big)  
-\frac{f(t)}{4\cdot 5!}F_5^2 \Big], \label{eq:totalaction}
\end{eqnarray}
where 
\begin{eqnarray}
f(t) &=& 1+t+\frac{1}{2}t^2, \\
g(t) &=& m^2t^2  + {\mathcal O}(t^4) .
\end{eqnarray}
Here we set $\alpha'=1$.
$f(t)$ describes  the coupling of the tachyon to the 5-form field strength $F_5$.
As noted in the previous subsection, $F_5$ is not constrained to be self-dual.
$g(t)$ is the  tachyon potential and $m^2(=-2/\alpha')$ is the tachyon mass squared.

Classical solution of $N_c$ stack of $D$-branes  in this theory has been studied in 
\cite{Klebanov:1998yya}
\cite{Minahan:1998tm}.
They  have found the behaviors of the solution approximately in the UV and IR regions 
using the following ansatz,
\begin{eqnarray}
ds^2=e^{\frac{1}{2}\phi}\Big( 
e^{\frac{1}{2}\xi-5\eta}d\rho^2+e^{-\frac{1}{2}\xi}\eta^{\mu\nu}dx_\mu dx_\nu
+e^{\frac{1}{2}\xi-\eta}d\Omega^2_5
\Big), \label{eq:metricansatz}\\
C_{4+ \mu\nu\rho\sigma}= \epsilon_{\mu\nu\rho\sigma}  {C(\rho)}, \label{eq:rransatz}
\end{eqnarray}
where $\rho$ is related to the radial direction transverse to the 3-brane.
For the 4-form R-R potential $C_4$, $C_{0123}(\rho)$ has been taken to be  the only non-vanishing
component. Furthermore, all fields have been assumed to be  functions of $\rho$.
We give the forms of the solution in each region in order.


\subsubsection{UV asymptotic solution}
A solution valid in the UV region, $\rho \ll 1$, is given by
\cite{Minahan:1998tm},
\begin{eqnarray}
&& \phi = \ln(2^{15}Q^{-1}) -2 \log y 
,\\
&& \xi = \ln(2Q) -y +\frac{1}{y}
,  \\
&& \eta = \ln 2 -\frac{1}{2}y+\frac{1}{y} 
, \\
&& t= -1 
,
\end{eqnarray}
where $\rho/\rho_0=e^{-y}$.
$\rho_0$ is an integration constant. 
$Q$ is the total $D3$ brane charge which is proportional to $N_c$.
An expansion which has higher order terms in $1/y$ is given in 
\cite{Klebanov:1998yy}, but we focus on the above solution in this paper.
In the Einstein frame, we can see that the metric is asymptotically $AdS_5 \times S^5$
in the limit $y \rightarrow \infty$.
Furthermore, we can find that the gauge coupling dependence on $\rho$ as 
\begin{eqnarray}
\frac{1}{g^2_{YM}}= e^{-\phi} \propto Q \Big( \ln  \Big( \frac{\rho}{\rho_0} \Big) \Big)^2. \label{eq:ymcoupling}
\end{eqnarray}
We can find a running coupling as the  logarithmic behavior of the dilaton background.
$\rho_0$ can be considered as the gauge theory length scale.


\subsubsection{IR asymptotic solution}

In the IR limit $\rho \rightarrow \infty$, 
it has been discussed in \cite{Minahan:1998tm} that the solution behaves as\footnote{
In \cite{Grena:2000xw} , it has been shown that the UV solution can be connected to the IR solution.} 
\begin{eqnarray}
&& \phi \simeq \phi_1 \rho +\phi_0 , \label{eq:cnfirphi} \\
&& \xi \simeq \xi_1 \rho +\xi_0 ,\label{eq:cnfirxi}\\
&& \eta \simeq \eta_1 \rho +\eta_0, \label{eq:cnfireta}\\
&& t \simeq t_1 \rho +t_0 ,\label{eq:cnfirtachyon},
\end{eqnarray}
with the following conditions on integration constants $\phi_i$, $\xi_i$, $\eta_i$ and $t_i$ $(i=0,1)$,
\begin{eqnarray}
&& \phi_1, \xi_1, \eta_1 > 0, \quad 5\eta_1-\frac{1}{2}\phi_1-\frac{1}{2}\xi_1 > 0, \\
&& \frac{1}{2}\phi_1^2+\frac{1}{2}\xi_1^2-5\eta_1^2+\frac{1}{4}t_1^2=0.
\end{eqnarray}
These conditions are necessary for the solution to satisfy the equations of motion derived from the type 0B action 
(\ref{eq:totalaction}).

In \cite{Minahan:1998tm}, 
it has been argued that this IR solution can lead to a linear quark potential by 
calculating the Wilson loop.
The linear quark potential requires an additional condition on the constants :
\begin{eqnarray}
\phi_1 \geq \xi_1. \label{eq:lqcondi}
\end{eqnarray} 
As we will see later, the condition $\phi_1=\xi_1$ is same as the one for
the linear confinement behavior of highly excited spin-1 meson masses.


\subsection{$5D$ action}
Let us present the low energy effective action on $D9$ brane pairs at quadratic order in gauge fields, 
\begin{eqnarray}
&& S =-\int d^{10}x {\rm Tr}\Big[ 
 e^{-\phi} V(TT^\dag)  
  \sqrt{-{\rm det}{\tilde{g}_{MN}}} 
 \Big( 1+ 
  \frac{1}{4}\tilde{g}^{MN}\tilde{g}^{KL} \sum_{i=L,R}F^i_{MK}F^i_{NL}
 \Big)
\Big], \hspace{40pt} \label{eq:dbiaction}
\end{eqnarray} 
where 
\begin{eqnarray}
&& \tilde{g}_{MN}=g_{MN}+
\frac{1}{2}  (D_M T)^\dag (D_N T) + \frac{1}{2} (D_N T)^\dag (D_M T),\\
&&  F^{L,R}_{MN} = \partial_M A^{L,R}_N -\partial_N A^{L,R}_N,\\
&&  D_M T = \partial_M T +i T A_M^L -iA_M^R T.
\end{eqnarray}
$T$ is the open string tachyon and $V(TT^\dag)$ is its potential.
$A^{L,R}_M$ are the world-volume gauge fields.
Note that $T$ transforms in the bifundamental representation 
of the $U(N_f)_L \times U(N_f)_R$ flavor symmetry group.

We  only consider quadratic terms in the gauge fields because it is enough to find the mass spectrum.
At this level, the action is just the sum of $N_f^2$ copies of the abelian action.

The form of the action coincides with that of the world volume DBI action 
on brane-antibrane pairs proposed in 
\cite{Sen:2004nf}
\cite{Sen:2003tm}
\cite{Garousi:2004rd}, expanded in terms of 
the gauge fields.
In type 0 string theory, 
the DBI action on $D$-branes has the coupling of the closed string tachyon
due to the existence of the tachyon tad pole on $D$-branes
\cite{Klebanov:1998yy}
\cite{Garousi:1999fu}.
However, it has  been argued that the closed string tachyon may be stabilized due to 
the coupling of the tachyon and R-R field strength
\cite{Klebanov:1998yya}
\cite{Klebanov:1999um}.
This means that the tachyon reaches the minimum of the its potential and has a fixed constant
in the background introduced before.
Thus,  the coupling of the closed string tachyon may not affect the 
form of the effective action, 
at least in the low energy approximation.

In order to obtain the $5D$ effective action, we make the following assumptions : 
(1) $T$ depends only $\rho$.  
(2) $T = \tau(\rho)  \times {\bf 1}$, where ${\bf 1}$ is 
the $N_f\times N_f$ identity matrix. 
(3) The tachyon potential has the form 
\begin{eqnarray}
V(\tau) =e^{-\frac{1}{2}\tau^2}. \label{eq:otachyonpote}
\end{eqnarray} 
(4) Gauge fields $A^{L,R}_M$ are functions of $x$ and $\rho$.
(5) $A^R_\alpha=A^L_\alpha=0$, where $\alpha$ denotes the coordinates of $S^5$.

The condition (2) means that 
the tachyon expectation values do not depend on the flavor degrees of freedom.
(3) is necessary for the linear confinement behavior of highly excited meson masses.
We impose (4) and (5)  to make  the gauge fields  not to depend on the coordinate of $S^5$.

Assuming these, 
we can obtain the $5D$ effective action by the dimensional reduction.
Choosing a gauge $A^L_u=A^R_u=0$, 
we obtain
\begin{eqnarray}
S &=& -\int d^4xdu \Big[ V(\tau)e^{\frac{1}{4}\phi+\frac{5}{4}\xi-\frac{5}{2}\eta}
g^2_{xx}\sqrt{g_{uu}+(\partial_u \tau)^2} \nonumber\\
&& 
+\frac{1}{4}F(u)\eta^{\mu\nu}\eta^{\rho\sigma}\Big( V_{\mu\rho}V_{\nu\sigma}
+A_{\mu\rho}A_{\nu\sigma} \Big) 
 + \frac{1}{2}G(u)
\eta^{\mu\nu} \Big( V_{\mu u}V_{\nu u}  + A_{\mu u }A_{\nu u} \Big)
    \nonumber\\
    && \qquad
    +H(u) 
\eta^{\mu\nu}A_\mu A_\nu \Big],
\label{eq:dbiexp}
\end{eqnarray}
where we define
\begin{eqnarray}
V_M = \frac{A^L_M +A^R_M}{2} , \quad A_M = \frac{A^L_M-A^R_M}{2}.
\end{eqnarray}
The coefficients are given by 
\begin{eqnarray}
F (u) & = & V(\tau)e^{\frac{1}{4}\phi+\frac{5}{4}\xi-\frac{5}{2}\eta}
(g_{uu}+(\partial_u \tau)^2)^{\frac{1}{2}}, 
\label{eq:coef} \\
G (u) &=& V(\tau)e^{\frac{1}{4}\phi+\frac{5}{4}\xi-\frac{5}{2}\eta} g_{xx}
(g_{uu}+(\partial_u \tau)^2)^{-\frac{1}{2}} , 
\label{eq:coeg} \\
H(u) &=& V(\tau)e^{\frac{1}{4}\phi+\frac{5}{4}\xi-\frac{5}{2}\eta}g_{xx}
(g_{uu}+(\partial_u \tau)^2)^{\frac{1}{2}} \tau^2 , 
\label{eq:coeh}
\end{eqnarray}


\section{Chiral symmetry breaking}
The open string tachyons, in general,  transform in the bifundamental representation of the flavor group, 
and couple on the boundary to  scalar and pseudo-scalar bilinears of quark fields. 
It has been suggested in \cite{Casero:2007ae} that
tachyon condensation on brane-antibrane system describes the physics of the chiral symmetry breaking. 
The tachyon expectation values, which is
 determined by the classical solution of  the tachyon equation of motion derived from the above action,
  are related to the quark masses and the chiral condensate.\footnote{There are other attempts to incorporate the quark mass 
in holographic dual models \cite{Hashimoto:2008sr}\cite{Aharony:2008an}} 
In this section, we will see how these parameters appear  in our model.

Let us set the $A^L=A^R=0$,
we obtain 
\begin{eqnarray}
S=-\int d^4xdu \hspace{2mm}V(\tau)e^{\frac{1}{4}\phi+\frac{5}{4}\xi-\frac{5}{2}\eta}
g^2_{xx}\sqrt{g_{uu}+(\partial_u \tau)^2}. \label{eq:5ddbi}
\end{eqnarray}
The equation of motion for the tachyon field $\tau$ is given by 
\begin{eqnarray}
\partial_u \Big( 
V(\tau)
\frac{e^{\frac{1}{4}\phi+\frac{5}{4}\xi-\frac{5}{2}\eta}g_{xx}^2}{\sqrt{g_{uu}+(\partial_u \tau)^2}} \partial_u \tau
\Big)
- V'(\tau) e^{\frac{1}{4}\phi+\frac{5}{4}\xi-\frac{5}{2}\eta}g_{xx}^2 \sqrt{g_{uu}+(\partial_u \tau)^2}=0,
\label{eq:ostachyoneom}
\end{eqnarray} 
where the prime denote differentiation with respect to $\tau$.

Let us first consider the UV region.
In the UV region, we expect that value  of the tachyon becomes small.
Thus, we assume that the tachyon equation is well described by linear terms of $\tau$.
In this assumption, we can approximate the equation as, 
\begin{eqnarray}
\partial_y^2 \tau+  \frac{y-2}{y}  \partial_y \tau
+  \frac{4(2y-9)}{y^2} 
\tau=0.
\end{eqnarray}
The solution of this equation 
in the limit $y \rightarrow \infty$ behaves as\footnote{
We can solve this equation as an asymptotic expansion,
\begin{eqnarray}
&& \tau(y) = m_q \Big[ y^{-8} + \sum_{n=1}^\infty 
\frac{1}{n!}\prod_{m=1}^n [(m+7)(m+10)+36] y^{-n-8} \Big]\\
&& \qquad +  \sigma   e^{-y} \Big[ y^{10} + \sum_{n=1}^\infty 
\frac{1}{n!}\prod_{m=1}^n [36-(m-8)(m-11)] y^{10-n} \Big].
\end{eqnarray}}  
\begin{eqnarray}
\tau \sim  m_q + \sigma e^{-y} , \label{eq:ostachyonuvsol}
\end{eqnarray}
where $m_q$ and $\sigma$ are integration constants.
We now have two independent solutions and the  constants associated with the solutions.
Since $\tau$ is considered to be  dual to the quark bilinear, 
these two integration constants can be related to the quark mass and the chiral condensate.
This identification can be done by looking at the behavior of the solutions in the UV limit.
In the UV limit $\rho \rightarrow \infty$, the tachyon solution behaves as $
\tau(y) \sim m_q$.
Thus, 
the constant $m_q$ may be  related to quark mass, and 
the constant $\sigma$ 
 may be related to the chiral condensate.

In the IR region, value of  the open string tachyon is expected to be large
\cite{Casero:2007ae}.
Thus, we should consider nonlinear equation. 
It seems to be   difficult to find the general solution.  
However, we do not need to solve the equation of motion 
in order to find the behavior of linear confinement behavior. 
We only assume that the tachyon expectation value becomes large in the IR background.


\section{Meson mass spectrum}
Now let us turn to the meson sector.
In particular, we focus on finding the linear confinement behavior for 
highly excited vector and axial meson masses.
In \cite{Casero:2007ae},
a framework for obtaining such a behavior has been formulated. 
The general conditions necessary for the behavior
have been discussed.
In this section, we show that 
one of the condition naturally comes from 
the linear potential behavior for quark confinement.
We also show that other conditions can be imposed in the type 0 IR background.


\subsection{Vector fields}
Let us consider the vector field $V_\mu$. 
We assume that $V_\mu$ can be expanded in terms of the complete sets $\{ \psi_n (u) \}$ as
\begin{eqnarray}
V_\mu (x, u) = \sum_n {\mathcal V}^{n}_\mu (x) \psi_n(u), \label{eq:vexp}
\end{eqnarray}
with an appropriate normalization condition which will be specified below.
The equations of motion for $V_\mu (x,u)$ can be solved by choosing 
$\psi_n$ as the eigenfunctions satisfying 
\begin{eqnarray}
-\partial_u \Big( 
G(u) \partial_u \psi_n \Big) 
-F(u) m^2_n \psi_n =0, \label{eq:vectoreq}
\end{eqnarray}
with the normalization condition of $\psi_n$,
\begin{eqnarray}
\int du F(u) \psi_n \psi_m = \delta_{nm}.
\end{eqnarray}
From these, we can obtain
\begin{eqnarray}
S= -\int d^4x \sum_n \Big[
\frac{1}{4} {\mathcal V}^{(n)}_{\mu\nu} {\mathcal V}^{(m)\mu\nu} +
\frac{1}{2} m^2_n {\mathcal V}^{(n)}_{\mu} {\mathcal V}^{(n)\mu}
\Big].
\end{eqnarray}

Let us study the vector meson mass spectrum.
The spectrum of the highly excited meson masses depends on the background in the IR region.
A general argument in \cite{Casero:2007ae}
states that the conditions for linear confinement in the IR background are
(1) $g_{xx} \rightarrow const$, (2) $\sqrt{g_{uu}+(\partial_u \tau)^2}
\sim \partial_u \tau$, (3) $e^{-\tilde{\phi}}\sim V(\tau)$.\footnote{
$e^{-\tilde{\phi}}$ is defined as, 
\begin{eqnarray}
e^{-\tilde{\phi}}g^2_{xx}\sqrt{\tilde{g}_{uu}}=\int d\Omega_5 e^{-\phi}V(\tau)\sqrt{\rm{det} \tilde{g}},
\end{eqnarray}
where $\tilde{g}_{MN}=g_{MN}+\partial_M \tau \partial_N \tau$.}
The condition (2) implies $\tilde{g}_{uu} =g_{uu}+(\partial_u \tau)^2 \sim (\partial_u \tau)^2$.
Thus, the radial variable in the metric is proportional to  $\tau$. 
The Gaussian behavior of the tachyon potential is essential for the linear confinement behavior, as in 
\cite{Karch:2006pv}.
Here, we would like to impose appropriate conditions on the IR background.
In our case, it turns out that the conditions are
(1) $\phi_1=\xi_1$, (3)$\frac{\phi_1}{4}+\frac{5\xi_1}{4}-\frac{5\eta_1}{2}=0$.
(1) is a special case of the condition for the linear quark potential (\ref{eq:lqcondi}).
We can impose a constraint on the parameters to satisfy (3).
From (1) and (3), 
we now have 
\begin{eqnarray}
g_{\rho\rho}=e^{(\frac{1}{2}\phi_1 +\frac{1}{2}\xi-5\eta_1) \rho}
 =e^{(\phi_1 -5\eta_1 )\rho} ,
\end{eqnarray}
with $\phi_1 < 5\eta_1$.
Therefore,   in the IR limit $\rho \rightarrow \infty$, 
we can see that (2) is automatically satisfied in our background.
Therefore, we can reproduce the linear confinement behavior for highly excited vector meson masses.


\subsection{Axial vector fields}
The axial vector field fluctuation $A_\mu$ can be split in a transverse and a longitudinal part, 
$A_\mu=A^{\bot}_\mu +A^{\parallel}_\mu$, with $\partial^\mu A^{\bot}_\mu=0$.
Here, we  consider the transverse part, corresponding to the axial vector meson excitation.
We expand $A_\mu^\bot$ as,  
\begin{eqnarray}
A^{\bot}_\mu (x,u) =\sum_n {\mathcal A}_\mu^{\bot (n)} (x) \phi_n (u).
\end{eqnarray}
We choose $\phi_n(x)$ as 
\begin{eqnarray}
-\partial_u \Big( 
G(u) \partial_u \phi_n \Big) 
-F(u) \lambda_n \phi_n + H(u)\phi_n =0,
\end{eqnarray}
with the normalization conditions
\begin{eqnarray}
\int du F(u) \phi_n \phi_m =\delta_{nm}.
\end{eqnarray}
We obtain, 
\begin{eqnarray}
S=-\int d^4x \sum_n \Big[
\frac{1}{4} {\mathcal A}^\bot_{(n)\mu\nu}{\mathcal A}^{\bot\mu\nu}_{(n)} 
+ \frac{1}{2}
H(u) {\mathcal A}^\bot_{(n)\mu}{\mathcal A}^{\bot\mu}_{(n)}
\Big].
\end{eqnarray}
We can find that the equation of motion has the same form as  the vector case
under the conditions discussed above. 
Therefore, in this case, 
we also have  the linear confinement behavior, $m_n^2 \propto n$, as in the vector case.



\section{Conclusions}
In this paper, we have studied a holographic dual model which has a type 0 string theory 
set up.
The set up we have considered is a $D$-brane system of 
the electric $D3$ and the probe $D9_-/\overline{D9}_-$ branes. 
We have studied the $5D$ effective action 
on the $D3$ background.
In particular, 
we have calculated the open string tachyon solution in the UV region. 
We have also shown that the vector and the axial vector meson mass spectrum 
have the behavior of linear confinement.

One interesting feature of type 0 string theory is that 
the Yang-Mills coupling has the logarithmic dependence on the coordinate transverse to $D3$-branes,  
(\ref{eq:ymcoupling}). 
Since this behavior appears in the UV background, 
this is expected to be related to the  high energy dynamics of the gauge theory. 
In QCD, there exist logarithmic scaling violations. 
These are due to the high energy processes.
It is interesting to investigate high energy scatterings 
and how we can see the scaling violations 
in type 0 string theory.

\section*{Acknowledgments} 
I am grateful to Y.~Baba, N.~Hatano N.~Ishibashi and Y.~Satoh 
 for useful discussions.
I would like to thank N.~Ishibashi for carefully reading  manuscript.
This work is supported by Grant-in-Aid for JSPS Research Fellows.



\begin{thebibliography}{99}
\bibitem{Maldacena:1997re}
  J.~M.~Maldacena,
  ``The large N limit of superconformal field theories and supergravity,''
  Adv.\ Theor.\ Math.\ Phys.\  {\bf 2}, 231 (1998)
  [Int.\ J.\ Theor.\ Phys.\  {\bf 38}, 1113 (1999)]
  [arXiv:hep-th/9711200].
  S.~S.~Gubser, I.~R.~Klebanov and A.~M.~Polyakov,
  ``Gauge theory correlators from non-critical string theory,''
  Phys.\ Lett.\  B {\bf 428}, 105 (1998)
  [arXiv:hep-th/9802109].
  E.~Witten,
  ``Anti-de Sitter space, thermal phase transition, and confinement in  gauge theories,''
  Adv.\ Theor.\ Math.\ Phys.\  {\bf 2}, 505 (1998)
  [arXiv:hep-th/9803131].

\bibitem{Karch:2002sh}
  A.~Karch and E.~Katz,
  ``Adding flavor to AdS/CFT,''
  JHEP {\bf 0206}, 043 (2002)
  [arXiv:hep-th/0205236].

\bibitem{Kruczenski:2003uq}
  M.~Kruczenski, D.~Mateos, R.~C.~Myers and D.~J.~Winters,
  JHEP {\bf 0405}, 041 (2004)
  [arXiv:hep-th/0311270].
  
\bibitem{Sakai:2004cn}
  T.~Sakai and S.~Sugimoto,
  ``Low energy hadron physics in holographic QCD,''
  Prog.\ Theor.\ Phys.\  {\bf 113}, 843 (2005)
  [arXiv:hep-th/0412141].

  
\bibitem{Erdmenger:2007cm}
  J.~Erdmenger, N.~Evans, I.~Kirsch and E.~Threlfall,
  ``Mesons in Gauge/Gravity Duals - A Review,''
  arXiv:0711.4467 [hep-th].
  
  
\bibitem{Erlich:2005qh}
  J.~Erlich, E.~Katz, D.~T.~Son and M.~A.~Stephanov,
  ``QCD and a holographic model of hadrons,''
  Phys.\ Rev.\ Lett.\  {\bf 95}, 261602 (2005)
  [arXiv:hep-ph/0501128].
  
\bibitem{Da Rold:2005zs}
  L.~Da Rold and A.~Pomarol,
  ``Chiral symmetry breaking from five dimensional spaces,''
  Nucl.\ Phys.\  B {\bf 721}, 79 (2005)
  [arXiv:hep-ph/0501218].



\bibitem{Polchinski:2001tt}
  J.~Polchinski and M.~J.~Strassler,
  ``Hard scattering and gauge/string duality,''
  Phys.\ Rev.\ Lett.\  {\bf 88}, 031601 (2002)
  [arXiv:hep-th/0109174].

\bibitem{Karch:2006pv}
  A.~Karch, E.~Katz, D.~T.~Son and M.~A.~Stephanov,
  ``Linear confinement and AdS/QCD,''
  Phys.\ Rev.\  D {\bf 74}, 015005 (2006)
  [arXiv:hep-ph/0602229].


\bibitem{Dixon:1986iz}
  L.~J.~Dixon and J.~A.~Harvey,
  ``String Theories In Ten-Dimensions Without Space-Time Supersymmetry,''
  Nucl.\ Phys.\  B {\bf 274}, 93 (1986).
  N.~Seiberg and E.~Witten,
  ``Spin Structures In String Theory,''
  Nucl.\ Phys.\  B {\bf 276}, 272 (1986).
  
\bibitem{Polyakov:1998ju}
  A.~M.~Polyakov,
  ``The wall of the cave,''
  Int.\ J.\ Mod.\ Phys.\  A {\bf 14}, 645 (1999)
  [arXiv:hep-th/9809057].

\bibitem{Cotrone:2007gs}
  A.~L.~Cotrone,
  ``On the YM and QCD spectra from five dimensional strings,''
  arXiv:0707.1483 [hep-th].



\bibitem{Klebanov:1998yya}
  I.~R.~Klebanov and A.~A.~Tseytlin,
  ``D-branes and dual gauge theories in type 0 strings,''
  Nucl.\ Phys.\  B {\bf 546}, 155 (1999)
  [arXiv:hep-th/9811035].
  
  
\bibitem{Klebanov:1998yy}
  I.~R.~Klebanov and A.~A.~Tseytlin,
  `` Asymptotic freedom and infrared behavior in the type 0 string approach  to gauge theory,''
  Nucl.\ Phys.\  B {\bf 547}, 143 (1999)
  [arXiv:hep-th/9812089].
  
  
\bibitem{Minahan:1998tm}
  J.~A.~Minahan,
  ``Glueball mass spectra and other issues for supergravity duals of {QCD} models,''
  JHEP {\bf 9901}, 020 (1999)
  [arXiv:hep-th/9811156].
  J.~A.~Minahan,
  ``Asymptotic freedom and confinement from type 0 string theory,''
  JHEP {\bf 9904}, 007 (1999)
  [arXiv:hep-th/9902074].
  
\bibitem{Casero:2007ae}
  R.~Casero, E.~Kiritsis and A.~Paredes,
  ``Chiral symmetry breaking as open string tachyon condensation,''
  Nucl.\ Phys.\  B {\bf 787}, 98 (2007)
  [arXiv:hep-th/0702155].

  
\bibitem{Dhar:2007bz}
  A.~Dhar and P.~Nag,
  ``Sakai-Sugimoto model, Tachyon Condensation and Chiral symmetry Breaking,''
  JHEP {\bf 0801}, 055 (2008)
  [arXiv:0708.3233 [hep-th]].
  ``Tachyon condensation and quark mass in modified Sakai-Sugimoto model,''
  arXiv:0804.4807 [hep-th].



\bibitem{Bergman:2007pm}
  O.~Bergman, S.~Seki and J.~Sonnenschein,
  ``Quark mass and condensate in HQCD,''
  JHEP {\bf 0712}, 037 (2007)
  [arXiv:0708.2839 [hep-th]].
  
  
  \bibitem{Polchinski:1998}
  J.~Polchinski, 
  ``String Theory, Vol. 2'' 
  Cambridge University Press, 1998.
  
  
  
\bibitem{Bergman:1997rf}
  O.~Bergman and M.~R.~Gaberdiel,
  ``A non-supersymmetric open-string theory and S-duality,''
  Nucl.\ Phys.\  B {\bf 499}, 183 (1997)
  [arXiv:hep-th/9701137].
  
\bibitem{Costa:1999qx}
  M.~S.~Costa,
  ``Intersecting D-branes and black holes in type 0 string theory,''
  JHEP {\bf 9904}, 016 (1999)
  [arXiv:hep-th/9903128].




  
  
\bibitem{Grena:2000xw}
  R.~Grena, S.~Lelli, M.~Maggiore and A.~Rissone,
  ``Confinement, asymptotic freedom and renormalons in type 0 string duals,''
  JHEP {\bf 0007}, 005 (2000)
  [arXiv:hep-th/0005213].


  
  
 


  
  


  
\bibitem{Sen:2004nf}
  A.~Sen,
  ``Tachyon dynamics in open string theory,''
  Int.\ J.\ Mod.\ Phys.\  A {\bf 20}, 5513 (2005)
  [arXiv:hep-th/0410103].
  
\bibitem{Sen:2003tm}
  A.~Sen,
  ``Dirac-Born-Infeld action on the tachyon kink and vortex,''
  Phys.\ Rev.\  D {\bf 68}, 066008 (2003)
  [arXiv:hep-th/0303057].
  
\bibitem{Garousi:2004rd}
  M.~R.~Garousi,
  ``D-brane anti-D-brane effective action and brane interaction in open  string channel,''
  JHEP {\bf 0501}, 029 (2005)
  [arXiv:hep-th/0411222].



\bibitem{Garousi:1999fu}
  M.~R.~Garousi,
  ``String scattering from D-branes in type 0 theories,''
  Nucl.\ Phys.\  B {\bf 550}, 225 (1999)
  [arXiv:hep-th/9901085].





\bibitem{Klebanov:1999um}
  I.~R.~Klebanov,
  ``Tachyon stabilization in the AdS/CFT correspondence,''
  Phys.\ Lett.\  B {\bf 466}, 166 (1999)
  [arXiv:hep-th/9906220].


\bibitem{Hashimoto:2008sr}
  K.~Hashimoto, T.~Hirayama, F.~L.~Lin and H.~U.~Yee,
  ``Quark Mass Deformation of Holographic Massless QCD,''
  arXiv:0803.4192 [hep-th].



\bibitem{Aharony:2008an}
  O.~Aharony and D.~Kutasov,
  ``Holographic Duals of Long Open Strings,''
  arXiv:0803.3547 [hep-th].

  
  

  
 

  
  


  
    

\end{thebibliography}
\end{document}